\documentclass{article}

\usepackage{arxiv}

\usepackage[utf8]{inputenc} 
\usepackage[T1]{fontenc}    
\usepackage{hyperref}       
\usepackage{url}            
\usepackage{booktabs}       
\usepackage{amsfonts}       
\usepackage{nicefrac}       
\usepackage{microtype}      
\usepackage{lipsum}
\usepackage{graphicx}

\usepackage{algpseudocode}
 \usepackage{verbatim}
  \usepackage{amsmath, amssymb}
  \usepackage{tikz}
  \usepackage[inline]{enumitem}
  \usepackage{siunitx}
   \usepackage{float}
   \usepackage{fancyhdr}

   \def\x{{\mathbf x}}
   
   \renewcommand{\vec}[1]{\mathbf{#1}}
   \DeclareMathOperator\erf{erf}

\graphicspath{ {./images/} }

\title{Enhancement by postfiltering for speech and audio coding in ad-hoc sensor networks}

\author{
 Sneha Das \\
 Department of Signal Processing and Acoustics\\
  School of Electrical Engineering\\
  Aalto University\\
  02230 Espoo, Finland \\
  \texttt{sneha.das@aalto.fi} \\
   \And
 Tom B\"ackstr\"om \\
 Department of Signal Processing and Acoustics\\
  School of Electrical Engineering\\
  Aalto University\\
  02230 Espoo, Finland \\
  \texttt{tom.backstrom@aalto.fi} \\
}

\begin{document}
\maketitle
\begin{abstract}
  Enhancement algorithms for wireless acoustics sensor networks~(WASNs) are indispensable with the increasing availability and usage of connected devices with microphones. Conventional spatial filtering approaches for enhancement in WASNs approximate quantization noise with an additive Gaussian distribution, which limits performance due to the non-linear nature of quantization noise at lower bitrates. In this work, we propose a postfilter for enhancement based on Bayesian statistics to obtain a multidevice signal estimate, which explicitly models the quantization noise. Our experiments using PSNR, PESQ and MUSHRA scores demonstrate that the proposed postfilter can be used to enhance signal quality in ad-hoc sensor networks. ~\footnote{Audio samples at \href{github.com/snehadas/postfilterASN/tree/master/sound_samples}{github.com/snehadas/postfilterASN/tree/master/sound\_samples}}
\end{abstract}


\thispagestyle{fancy}
\lhead{Forthcoming, \\The Journal of the Acoustical Society of America Express Letters~(JASA-EL).}

\section{Introduction}
The emergence of connected and portable devices like smartphones, and the rising popularity
of voice user-interfaces and devices equipped with microphones, enable the necessary infrastructure for ad-hoc
wireless acoustic sensor networks~(WASNs). The dense, ad-hoc positioning and collaboration in a WASN leads to efficient sampling of the acoustic space, thereby gaining higher quality signal estimates compared to single-channel estimates~\cite{bertrand2011applications}.
Typical applications of ad-hoc WASNs use microphones on low-resource devices, such that we need low-complexity methods and which use bandwidth efficiently to compress and transmit the acoustic signals. This involves quantization at the encoder, whereby the received signal at the decoder is usually degraded by quantization noise~\cite{pradhan2003distributed, dragotti2009distributed, backstrom2017celp,backstrom2017fast,baeckstroem2016coding}.

Past works on WASN often overlook the variability in maximum capacity of sensors~\cite{zahedi2015coding}. However, rate-constrained spatial filtering like beamforming and multichannel Wiener filtering have been used in binaural hearing aids~(HAs)~\cite{roy2008rate, srinivasan2009rate, srinivasan2009analyzing, doclo2009reduced, dragotti2009distributed}. A study on  rate-constrained optimal beamforming showed the advantage of using spatially separated microphones in HAs, although the method assumes that the joint statistics of signals are available at the processing nodes~\cite{roy2008rate}. Subsequently, sub-optimal strategies for noise reduction which do not use the joint statistics at the nodes have been proposed~\cite{roy2008rate, srinivasan2009rate, srinivasan2009analyzing, doclo2009reduced, amini1}.
While the above methods are effective in reducing noise, they are either limited to, or are most efficient with two nodes~(HAs) only. In a recent work on multi-node WASN, a linearly-constrained minimum variance beamformer was used to optimize rate allocation and sensor selection over nodes, based on spatial location and frequency content~\cite{amini2019rate,zhang2017microphone}. However, due to the dynamic nature of an ad-hoc WASN, information about sensor distribution, location, number of target and interference sources may be either unavailable, or their exchange between nodes further adds to the bandwidth consumption and communication complexity. Further, the above methods assume an additive quantization noise model, which is accurate only at higher bitrates. Lastly, while all the above methods are optimized on Wyner-Ziv coding, their suitability in combination with existing speech and audio coding has not been demonstrated yet. Their performance in single-channel mode can therefore not compete with conventional single-channel codecs.

\begin{figure}[!t]
\centering
\resizebox{0.45\columnwidth}{!}{
\begin{tikzpicture}[scale=0.99, transform shape]

\node at (1.2,1.5) (speaker){\includegraphics[width=0.7cm]{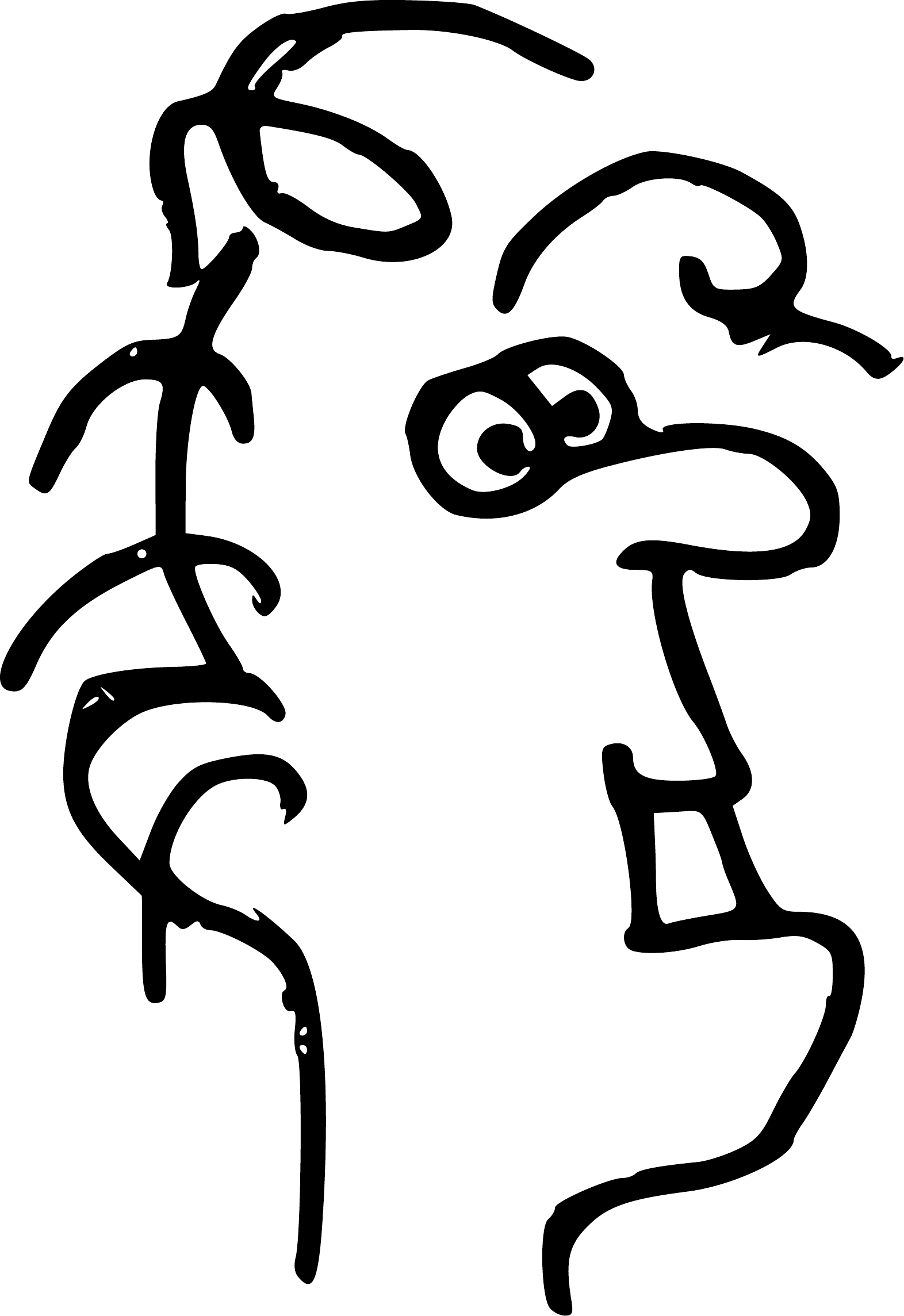}};
\node[align=center] at (1.2, 0.4){Speech-\\source};
\draw[thick,domain=-40:40]  plot({1.6+0.2*cos(\x)}, {1.3+0.2*sin(\x)});
\draw[thick,domain=-40:40]  plot({1.6+0.4*cos(\x)}, {1.3+0.4*sin(\x)});
\draw[thick,domain=-40:40]  plot({1.6+0.6*cos(\x)}, {1.3+0.6*sin(\x)});

\node at (3.2,1.5) (micCh1){\includegraphics[width=0.3cm]{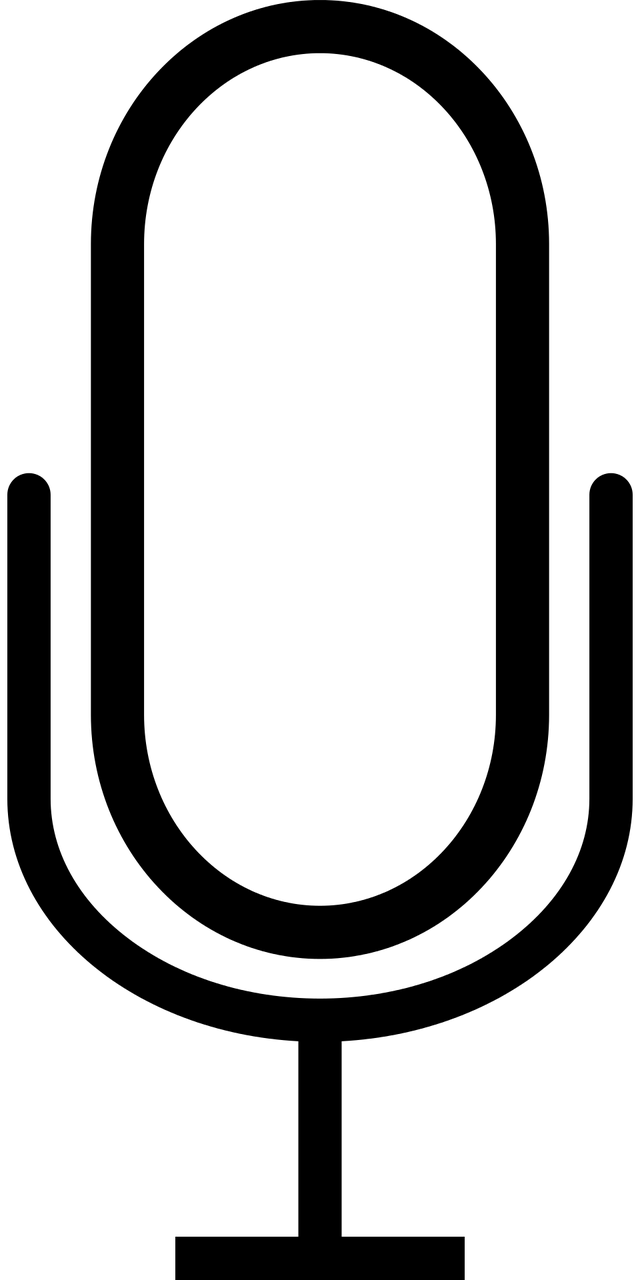}};
\node[align=center] at (3.2, 0.8){Device A};

\node at (4,2.3) (micCh2){\includegraphics[width=0.3cm]{figures/mic.png}};
\node[align=center] at (4.2, 1.6){Device B};

\node at (8,1.5) (noise){\includegraphics[width=0.4cm, angle=180]{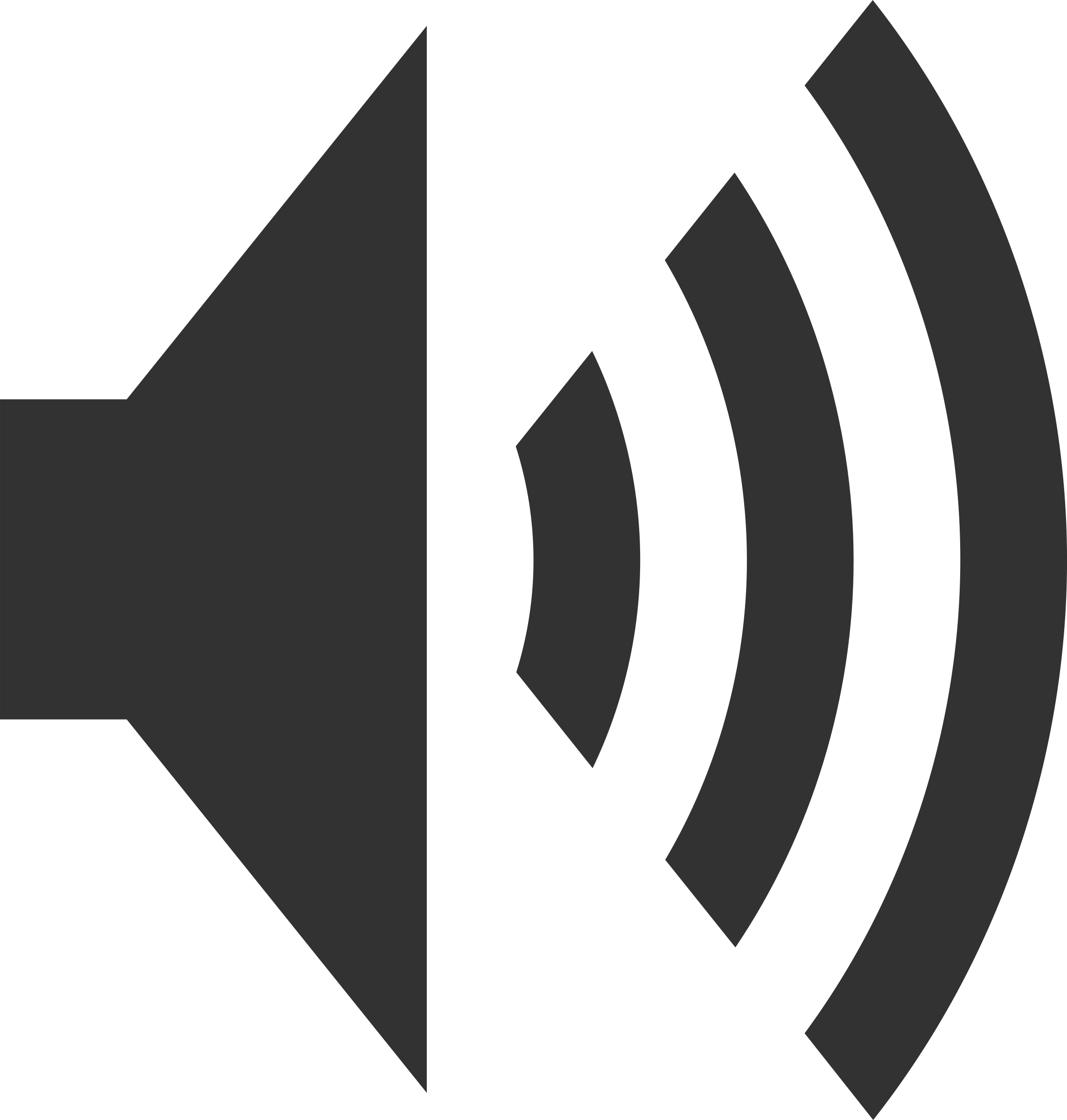}};
\node[align=center] at (6.8, 1.5){Noise-\\sources};
\node at (6,2.3){\includegraphics[width=0.4cm, angle=180]{figures/speaker.png}};
\node at (7,0.4){\includegraphics[width=0.4cm, angle=180]{figures/speaker.png}};
\draw[black](0.5, 2.7)rectangle(8.5, 0.0);



\end{tikzpicture}}
\caption{\footnotesize{Distribution of microphones in the ad-hoc acoustic sensor network.}}

\label{dist_interfaces}

\end{figure}

In this paper, we propose a Bayesian postfilter for enhancement in ad-hoc WASNs, which explicitly models the quantization noise within the optimization framework of the filter, and can be applied on top of existing codecs with minimal modifications.
Thus, the main contribution of the current work is the postfilter which takes quantization into account through truncation, while retaining the conventional assumption of additive Gaussian background noise, thereby resulting in a truncated Gaussian representation of the clean speech distribution.
 To evaluate the proposed methodology, we place the necessary assumptions that the devices are dominantly degraded by either background noise and reverberation, or coding noise due to quantization, and each device operates at its maximum capacity.  In line with past works, we show that by distributing the total available bitrate between the two
sensors, the output gain of the WASN signal estimate
is higher than the output gain of a low input-SNR single sensor transmitting at
full bitrate~\cite{roy2008rate, srinivasan2009rate, srinivasan2009analyzing, doclo2009reduced}. In addition, we present the advantages of incorporating the exact quantization noise models within the optimization framework. In order to focus on the effect of the postfilter on quantization noise, we apply the proposed method on the output of a codec~\cite{backstrom2018dithered}, which is specifically designed to address multi-device coding. To the best of our knowledge, this is the first time a complete WASN system is evaluated with competitive performance also in a single-channel codec mode. Although we have not yet included models of spatial configuration of sensors, room impulse responses or multiple sources, we show that the proposed method already yields large output gains.

%

\lhead{}
 \section{Methodology}

 To focus on the novel aspects of the approach, we consider a simple WASN consisting of two devices with microphones: \begin{enumerate*}
 \item a low-resource device~A
 with high input SNR and \item a high-resource device~B with low input SNR,
 \end{enumerate*}
 as illustrated in Fig.~\ref{dist_interfaces}.
 An example
 application is a smartwatch that collaborates with a distant
 smart speaker.
 Let \(x(k, t), n(k, t)\) be the perceptual domain representations of the
 speech
 and noise signal, respectively, at the frequency bin \(k\) and time frame
 \(t\)~\cite{backstrom2017speech}; the perceptual domain representations are computed by dividing the frequency domain signals by the perceptual envelope obtained from the codec~\cite{backstrom2017speech}. These signals can be approximated by zero-mean Gaussian
 distributions with variances $\sigma_x^2$ and $\sigma_n^2$, whereby the random variables
 are correspondingly \(  X\sim\mathcal{N}(0, \sigma_{x}^{2}), \quad
   N\sim\mathcal{N}(0, \sigma_{n}^{2})\)~\cite{kim2008gaussianity}.
    Under the
   assumption of uncorrelated, additive background noise, the noisy signal \(y(k, t)=x(k, t)+ n(k, t)\) is Gaussian distributed with  \(Y\sim\mathcal{N}(0,\sigma_y^2)\),
  and variance \(\sigma_y^2 = \sigma_x^2+\sigma_n^2\)~\cite{kim2008gaussianity}. Our
  goal is to estimate the
   distribution of clean speech, conditioned over the noisy observation \(P(X\mid Y)\),
   in other words, the {\it posterior distribution}~\cite{sarkka2013bayesian}.
   We obtain estimates for every time-frequency bin, and shall omit the time and frequency subscripts in the rest of the section to aid readability. According to the Bayes rule, the posterior distribution can be written as:
     \begin{equation}\label{eq:2}
       P(X\mid Y) = \frac{P(X)P(Y\mid X)}{P(Y)}\propto P(X)P(Y\mid X),
     \end{equation}
       where \(P(X)\) and \(P(Y)\) are the {\it prior
       distributions} of the speech and observed signals and \(P(Y\mid X)\) is the conditional likelihood. However, our quantized observation, $y_q(k,t)$ of the noisy signal gives more evidence about \(X\); The true value of the noisy signal \(Y\) lies within the quantization bin limits,
       \(y(k, t)\in[l(k,t), u(k,t)]\) and the lower and upper bin limits for the
       quantization
       levels in a frame \(\{\vec{l}, \vec{u}\}\in\mathbb{R}^{K\times1}\) are
       obtained from the observed quantized spectrum of a frame
       \(\vec{y}_q\in  \mathbb{R}^{K\times1}\)~\cite{snehaLogMag}. Since the true noisy signal lies in the bounded field \(l(k,t) \leq Y\leq u(k,t)\), we compute the summation of the likelihood over the quantization bin limits to obtain the posterior distribution of speech,
 \begin{equation}
   \label{eq:3}
 P(X)_{(l\leq Y\leq u)}\propto P(X)\int_{l}^{u} P(Y\mid X)dy,
 \end{equation}
 where $\propto$ signifies equality up to a scaling factor. Eq.~\ref{eq:3} can be rewritten as the difference between cumulative
 distributions, \( P(X)_{(l\leq Y\leq u)} \propto P(X)(F(u)-F(l))\). The conditional likelihood can be represented as \(P(Y\mid X)\sim \mathcal{N}(x, \sigma_{n}^{2})\), thus
 resulting in the final equation
  for the posterior distribution,
  \begin{equation}\label{eq:4}
  P(X)_{(l\leq Y\leq u)}\propto P(X)\left[0.5\bigg\{\erf\left(\frac{u-x}{\sigma_n\sqrt{2}}\right)-\erf\left(\frac{l-x}{\sigma_n\sqrt{2}}\right)\bigg\}\right],
 \end{equation}
 where \(\erf(.)\) is the error function. Note that due to the use of the exact quantization bin limits, \( P(X)_{(l\leq Y\leq u)}\) corresponds to a {\it truncated Gaussian}~\cite{barr1999mean}. This is in contrast to past works, where the quantization noise is {\it approximated} by an additive Gaussian distribution, which is an accurate approximation only at high bitrates~\cite{amini2019rate}.

 \begin{figure}[!tb]
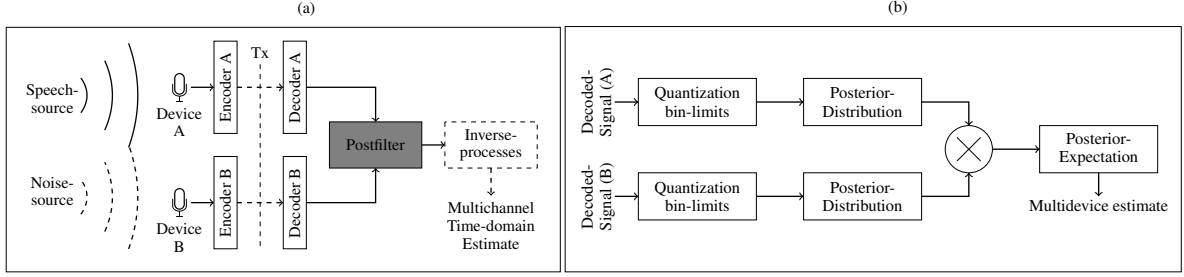

\centering
\resizebox{0.45\columnwidth}{!}{
\begin{tikzpicture}[scale=0.99, transform shape]
\draw[black](0, 0)rectangle(0.5, 2);
\node[align=center, rotate = 90]at(0.25, 1){Encoder A};

\draw[black](0, -2.5)rectangle(0.5, -0.5);
\node[align=center, rotate = 90]at(0.25, -1.5){Encoder B};

\draw[black](0+1.5, 0)rectangle(0.5+1.5, 2);
\node[align=center, rotate = 90]at(0.25+1.5, 1){Decoder A};

\draw[black](0+1.5, -2.5)rectangle(0.5+1.5, -0.5);
\node[align=center, rotate = 90]at(0.25+1.5, -1.5){Decoder B};

\draw[black, fill=gray](2.5, -0.75)rectangle(4.5, 0.25);
\node[align=center]at(3.5, -0.25){Postfilter};


\draw[black, dashed](4.5+0.5, -1.25+1.5)rectangle(6.5+0.5, -2.25+1.5);
\node[align=center]at(6, -1.75+1.5){Inverse-\\processes};

\node at (-0.75,1) (micCh1){\includegraphics[width=0.3cm]{figures/mic.png}};
\node[align=center] at (-0.75, 0.25){Device\\A};

\node at (-0.75,-1.5) (micCh1){\includegraphics[width=0.3cm]{figures/mic.png}};
\node[align=center] at (-0.75, -2.25){Device\\B};

\node[align=center] at (-3.5, 0.75){Speech-\\source};
\draw[scale=5.5,thick,domain=-20:20]  plot({-0.9+0.4*cos(\x)}, {0.15+0.2*sin(\x)});
\draw[scale=5.5,thick,domain=-20:20]  plot({-0.9+0.5*cos(\x)}, {0.15+0.4*sin(\x)});
\draw[scale=5.5,thick,domain=-20:20]  plot({-0.9+0.6*cos(\x)}, {0.15+0.6*sin(\x)});

\node[align=center] at (-3.5, -1.25){Noise-\\source};
\draw[scale=5.5,thick,domain=-20:20, dashed]  plot({-0.9+0.4*cos(\x)}, {-0.25+0.2*sin(\x)});
\draw[scale=5.5,thick,domain=-20:20, dashed]  plot({-0.9+0.5*cos(\x)}, {-0.25+0.4*sin(\x)});
\draw[scale=5.5,thick,domain=-20:20, dashed]  plot({-0.9+0.6*cos(\x)}, {-0.25+0.6*sin(\x)});

\draw[thick, ->](-0.5,1)--(0,1);
\draw[thick, dashed, ->](0.5,1)--(1.5,1);
\draw[thick, ->](-0.5,-1.5)--(0,-1.5);
\draw[thick, dashed, ->](0.5,-1.5)--(1.5,-1.5);

\draw[thick, ->](2, 1)--(3.5,1)--(3.5,0.25);
\draw[thick, ->](2, -1.5)--(3.5,-1.5)--(3.5,-0.75);

\draw[thick, ->](4.5, -0.25)--(5, -0.25);
\draw[thick, dashed, ->](6, -2.25+1.5)--(6, -2.85+1.5);
\node[align=center]at(6, -3.5+1.5){Multichannel\\Time-domain \\Estimate};

\draw[dashed](1, 1.5)--(1, -2.5);
\node[align=center]at(1, 1.75){Tx};
\draw[black](-4.5, -3)rectangle(7.5, 2.3);
\node[align=center]at(2, 2.7){(a)};

\end{tikzpicture}}\resizebox{0.50\columnwidth}{!}{
\begin{tikzpicture}[scale=0.99, transform shape]

\draw[black](0, 0)rectangle(2.5, 1);
\node[align=center]at(1.25, 0.5){Quantization\\ bin-limits};
\draw[black, thick,->](2.5, 0.5)--(3.5, 0.5);

\draw[black](0, -2)rectangle(2.5, -1);
\node[align=center]at(1.25, -1.5){Quantization\\ bin-limits};
\draw[black, thick,->](2.5, -1.5)--(3.5, -1.5);

\draw[black](3.5, 0)rectangle(6, 1);
\node[align=center]at(4.75, 0.5){Posterior-\\Distribution};
\draw[black, thick, ->](6, 0.5)--(7, 0.5)-|(7, 0);

\draw[black](3.5, -2)rectangle(6, -1);
\node[align=center]at(4.75, -1.5){Posterior-\\Distribution};
\draw[black, thick, ->](6, -1.5)--(7, -1.5)-|(7, -1);

\draw(7,-0.5)circle(0.5cm);
\draw[black](6.75,-0.75)--(7.25,-0.25);
\draw[black](6.75,-0.25)--(7.25,-0.75);

\draw[black](8.5, -1)rectangle(11, 0);
\node[align=center]at(9.75, -0.5){Posterior-\\Expectation};
\draw[thick, black, ->](7.5, -0.5)--(8.5, -0.5);

\draw[black, thick,->](-0.5, 0.5)--(0, 0.5);
\node[align=center, rotate=90]at(-0.85, 0.5){Decoded-\\Signal~(A)};

\draw[black, thick,->](-0.5, -1.5)--(0, -1.5);
\node[align=center, rotate=90]at(-0.85, -1.5){Decoded-\\Signal~(B)};

\draw[thick, black, ->](9.75,-1)--(9.75,-1.5);
\node[align=center]at(9.75, -1.7){Multidevice estimate};

\draw[black](-1.55, -3.1)rectangle(11.5, 2.1);
\node[align=center]at(5.5, 2.5){(b)};
\end{tikzpicture}}
\caption{Block diagrams showing (a)~the overall system structure with the location of the postfilter, and (b)~overview of the postfilter.}
\label{sys_blockDiagram}
\end{figure}

 From Eq.~\ref{eq:4}, the single channel posterior probability distribution function~(PDF) of the clean speech in spatial channel \(i\) is \(
   P_i(X)_{(l_i\leq Y_i\leq u_i)}
 \propto P_i(X)\left[0.5\bigg\{\erf\left(\frac{u_i-x}{\sigma_{n_i}\sqrt{2}}\right)-\erf\left(\frac{l_i-x}{\sigma_{n_i}\sqrt{2}}\right)\bigg\}\right]
 \). Here we assume that the speech and noise energies at each channel are
 estimated in a pre-processing stage, for example, using voice activity detection and minimum
 statistics~\cite{martin2001noise}. Additionally, in order to focus on the advantage of the proposed enhancement approach, we assumed that the time-delay between microphones with respect to the desired sources was known at the decoder, whereby the signals from the
 microphones were synchronized. We shall include time-delay estimation within the enhancement framework in future work.
 Based on our setup, the environmental degradation and the bitrate
  are different for the two channels. Hence, we can assume that \(N_i\sim\mathcal{N}(\mu_{n_i},\sigma_{n_i}^2)\) and the quantization-bin \(\{\vec{l}, \vec{u}\}_i\) offsets are uncorrelated and independent between the two channels. Therefore, when conditioned on $Y$, due to conditional independence between the channels, the joint posterior PDF of speech
 over the network can
 be represented as \(P(X)_Y \propto \prod_{i=1}^M P_i(X)_{(l_i\leq Y_i\leq u_i)}\), where \(M\)
 is the number of microphones
 in the WASN. The posterior PDF of speech in a two microphone network is thus:
 \begin{multline}
 \label{eq:5}
     P(X)_Y \propto \frac{e^{-{\frac{1}{2}}\sum_{i=1}^{2}\left({\frac{x-\mu_{s_i}}{\sigma_{s_i}}} \right)^2}}{8\pi\prod_{i=1}^{2}\sigma_{s_i}} \prod_{i=1}^{2}\erf\left(\frac{u_i-x}{\sigma_{n_i}\sqrt{2}}\right)+ \\\prod_{i=1}^{2}\erf\left(\frac{l_i-x}{\sigma_{n_i}\sqrt{2}}\right)-\sum_{i=1}^{2}\left[\erf\left(\frac{u_i-x}{\sigma_{n_i}\sqrt{2}}\right)\erf\left(\frac{l_{3-i}-x}{\sigma_{n_{3-i}}\sqrt{2}}\right)\right].
 \end{multline}
 We obtain the multidevice signal estimate
 \(\hat{x}_{\text{MC}}\), optimal in minimum mean
 squared error~(MMSE) sense~\cite{sarkka2013bayesian} by computing the expectation of the PDF obtained from Eq.~\ref{eq:5}.
 Due to the product of error-functions in Eq.~\ref{eq:5}, the expectation does not have a known analytical formulation. Therefore, we approximate the expectation of the PDF via numerical integration~\cite{mcleod1980generalized}; computing the Riemann sum using the midpoint rule over intervals \(n=200\) provided an approximate with sufficient accuracy in our experiments. Hence, the final equation is
 \begin{equation}
 \label{finalEq}
     \hat{x}_{\text{MC}} =
     E[X_{\text{MC}}]\approx\sum_{j=1}^{n}x_{j}P(X=x_j)_{Y},  \quad \quad\min\limits_{i=1, 2}\{l_i\}\leq x\leq\max\{u_i\}
 \end{equation}

 \begin{figure*}[!t]
   \centering
   \includegraphics[width=1.0\textwidth]{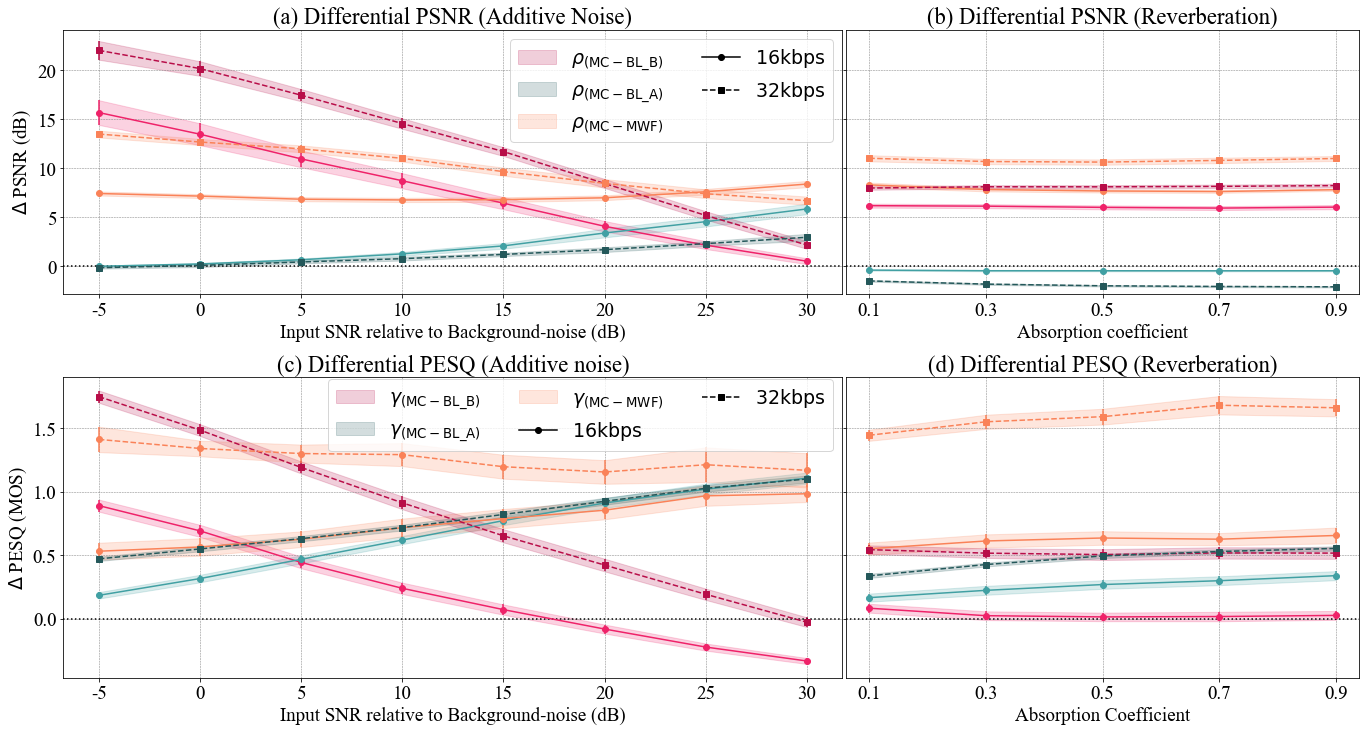}
 \caption{Illustration of differential PSNR and PESQ scores between the proposed multidevice estimate, and single-channel baseline and multichannel Wiener filter at \(R=\{16,\SI{32}{kbps}\}\) with 95\% confidence intervals.
 \(\rho_{(\text{MC}-\text{BL\_B})}\) and \(\gamma_{(\text{MC}-\text{BL\_B})}\) are the differential PSNR and PESQ of the proposed multidevice estimate with respect to single-channel estimate of device~B; \(\rho_{(\text{MC}-\text{BL\_A})}\) and \(\gamma_{(\text{MC}-\text{BL\_A})}\) are the differential scores of the multidevice estimate with respect to single-channel estimate of device~A; \(\rho_{(\text{MC}-\text{MWF})}\) and \(\gamma_{(\text{MC}-\text{MWF})}\) are the differential scores of the multidevice estimate with respect to the multichannel Wiener filter.}
 \label{error_bars}
 \end{figure*}
 The system block diagram is depicted in Fig.~\ref{sys_blockDiagram}~(a, b),
 where
 (a) is the overview of the entire system, from acoustic signal acquisition
 at the
 sensors to obtaining the time-domain estimate from multidevice signals.
 Note that the postfilter is placed at the fusion center, directly after the
 decoder, which provides the decoded perceptual domain signals to the
 postfilter. Fig.~\ref{sys_blockDiagram}~(b) shows the internal structure of
 the postfilter. After receiving the quantization bin limits from the decoded
 signals, we compute the  truncated Gaussian distribution for each channel,
 and then compute the joint posterior distribution as the product
 of the truncated distributions of the channels. The final point estimate,
 obtained as the expectation of the posterior distribution, yields the multidevice
 signal estimate.

  \section{Experiments and Results}
  To evaluate the performance of the proposed postfiltering approach, we
  determined the perceptual SNR~(PSNR) and PESQ
  scores~\cite{backstrom2017speech}, and conducted a subjective
  listening test using MUSHRA~\cite{series2014method, schoeffler2015towards}.
  We considered two categories of degradation:
  \begin{enumerate*}
  \item additive background noise,
  \item background noise with reverberation.
  \end{enumerate*}
  For the background noises, from the QUT dataset, we extracted the cafeteria scenario with babble noise~\cite{dean2010qut}.
  The clean speech samples were obtained from the
  test set of the TIMIT dataset~\cite{zue1990speech}. We encoded the noisy
  samples and applied the proposed postfilter to the decoded samples. Hence,
  the output signal is corrupted by both coding and environmental artefacts.
  To generate noisy speech with reverberation, we considered a room of
  dimensions \((7.5\times5\times2)\si{\meter^3}\), with one speech source at
  coordinates \((1, 2.5, 0.5)\si{\meter}\) and three noise sources placed at
  \((6.5, 2.85, 0.5)\si{\meter}, (3.5, 4.5, 0.5)\si{\meter}\) and
  \((6, 0, 0.5)\si{\meter}\). The locations of the near and distant
  microphones are, respectively, \((1.05,2.55, 0.5)\si{\meter}\) and
  \((2.25,2.85, 0.5)\si{\meter}\). An illustration of the setup is presented
  in Fig.~\ref{dist_interfaces}. The signals at the microphones for the
  described acoustic scenario were simulated using
  Pyroomacoustics~\cite{scheibler2018pyroomacoustics}.

    \begin{figure}[!t]
      \centering
      \includegraphics[width=0.69\columnwidth]{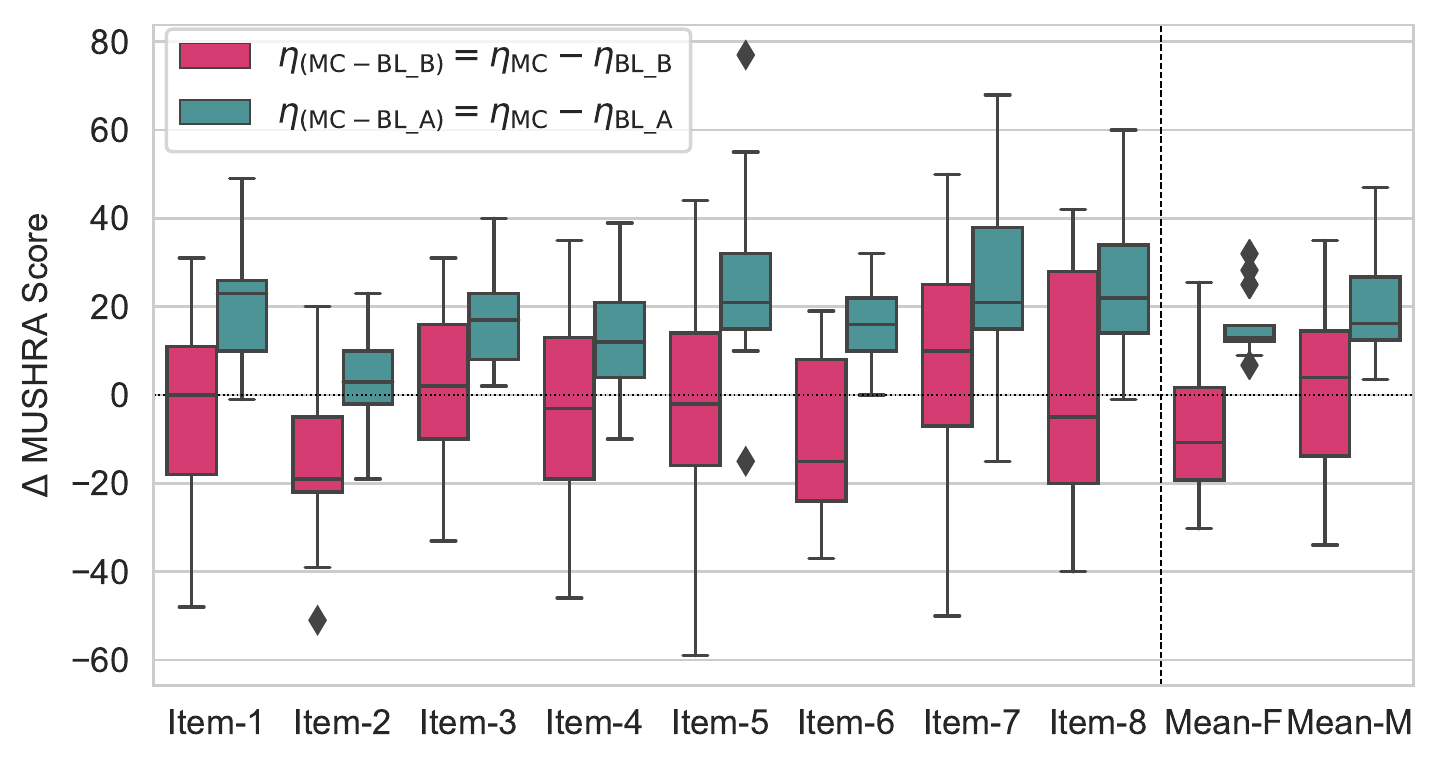}
      \caption{Distribution of \(\Delta\)MUSHRA points from the subjective listening test. \(\eta_{(\text{MC}-\text{BL\_B})}\) and \(\eta_{(\text{MC}-\text{BL\_A})}\) are the differential MUSHRA of multidevice estimate with respect to signal-channel estimates at device~B and device~A, respectively. Mean-F and Mean-M are the average differential scores over the female and males items, respectively.}
      \label{mushra}
    \end{figure}
  Let \(\rho\) and \(\gamma\) represent the PSNR and PESQ scores, respectively and the total bitrate is
  \(R\). The postfilter is applied on the output of a codec that is specifically suitable for multi-device coding~\cite{backstrom2018dithered}. For a fair evaluation, the single channel enhancement from Eq.~\ref{finalEq} are used as baselines. Furthermore, we employ the conventional multichannel Wiener Filter~(MWF) with diagonalized covariance matrix, to evaluate the advantage of the proposed method with respect to a conventionally accepted baseline~\cite{1025586}. The notations and their definitions are as
  follows: \begin{enumerate*}
  \item \(\hat{x}_{\text{MC}}\) is the multidevice estimate using device~A
  at the \(\text{bitrate}=\frac{1}{4}R\) and device~B at the
  \(\text{bitrate}=\frac{3}{4}R\); the PSNR and PESQ scores of the estimate are
  \(\rho_{\text{MC}}\), \(\gamma_{\text{MC}}\) respectively,
  \item \(\hat{x}_{\text{BL\_B}}\) is the baseline posterior estimate~(from Eq.~\ref{finalEq}) at distant device~B, encoding at full \(\text{bitrate}=R\); \(\rho_{\text{BL\_B}}\) and \( \gamma_{\text{BL\_B}} \) are the
  objective measures,
  \item \(\hat{x}_{\text{BL\_A}}\) is the baseline posterior estimate~(from Eq.~\ref{finalEq}) at device~A using \(\text{bitrate}=\frac{1}{4}R\), and \(\rho_{\text{BL\_A}}\),
  \( \gamma_{\text{BL\_A}} \) are the objective measures,
  \item \(\hat{x}_{\text{MWF}}\) is the multichannel Wiener filter using noisy signals from device~A and device~B, and \(\rho_{\text{MWF}}\),
  \( \gamma_{\text{MWF}} \) are the objective measures.
  \end{enumerate*} We show the advantage of the proposed postfilter over the baseline methods using differential PSNR and PESQ scores; their definitions are:
  \begin{enumerate*}
  \item\(\rho_{(\text{MC}-\text{BL\_B})}=\rho_{\text{MC}}-\rho_{\text{BL\_B}} \),
  \item\(\rho_{(\text{MC}-\text{BL\_A})}=\rho_{\text{MC}}-\rho_{\text{BL\_A}}  \),
  \item\(\rho_{(\text{MC}-\text{MWF})}=\rho_{\text{MC}}-\rho_{\text{MWF}}  \),
  \item\(\gamma_{(\text{MC}-\text{BL\_B})}=\gamma_{\text{MC}}-\gamma_{\text{BL\_B}}\),
  \item\(\gamma_{(\text{MC}-\text{BL\_A})}=\gamma_{\text{MC}}-\gamma_{\text{BL\_A}} \), \item\(\gamma_{(\text{MC}-\text{MWF})}=\gamma_{\text{MC}}-\gamma_{\text{MWF}} \).
  \end{enumerate*}

  The input SNR at device~A was fixed to \SI{40}{dB} and at device~B, and we
  used a range of input \(\text{SNRs} \in\{-5, 0, 5..., \SI{30}{dB}\}\).
  From the test set of the TIMIT dataset, we randomly selected 100 speech
  samples (50 male and 50 female) and tested the postfilter over all the
  combinations of the bitrates,
  \(R\in\{16, 24, 32, 48, 64, 80, 96\si{kbps}\}\), and the input SNRs for each
  speech sample. The objective results for the additive noise scenario are
  presented in Fig.~\ref{error_bars}~(a, c). \(\rho_{(\text{MC}-\text{BL\_A})}\), \(\rho_{(\text{MC}-\text{BL\_B})}\) and \(\rho_{(\text{MC}-\text{MWF})}\) are shown in Fig.~\ref{error_bars}~(a) for
  the listed SNRs and the total \(\text{bitrate}\in\{16, 32\si{kbps}\}\);
  We found that the PSNR  of the proposed method was better than all three baselines over all SNRs and bitrates. For \(\hat{x}_{\text{MC}}\) relative to the single-channel estimate \(\hat{x}_{\text{BL\_B}}\), the highest differential PSNR is  \(\rho_{(\text{MC}-\text{BL\_B})}\approx\SI{22.5}{dB}\).
  With respect to \(\hat{x}_{\text{BL\_A}}\), the highest \(\rho_{(\text{MC}-\text{BL\_A})}\approx\SI{6}{dB}\) is obtained at \(\SI{30}{dB}\) input SNR and \(\SI{16}{kbps}\). In addition, we observe that \(\rho_{(\text{MC}-\text{BL\_B})}\) decreases with the increase in the input SNR at device~B; also, it increases with an increase in total bitrate due to lower degradation from coding noise, specifically at device~A. In contrast, \(\rho_{(\text{MC}-\text{BL\_A})}\) increases with an increase in the input SNR at device~B but decreases with increase in the total bitrate. In terms of PESQ, the largest differential PESQ for \(\hat{x}_{\text{MC}}\) relative to \(\hat{x}_{\text{BL\_B}}\) is \(\gamma_{(\text{MC}-\text{BL\_B})}\approx\SI{1.8}{MOS}\), attained at \(\SI{-5}{dB}\) and \(\SI{32}{kbps}\). However, at \SI{16}{kbps} and above \SI{15}{dB} the negative
  \si{MOS} implied a decrease in quality. With respect to \(\hat{x}_{\text{BL\_A}}\), largest value is \(\gamma_{(\text{MC}-\text{BL\_B})}\approx\SI{1.1}{MOS}\) at \(\SI{30}{dB}\) input SNR at device~B. Furthermore, the variations of \(\gamma_{(\text{MC}-\text{BL\_A})}\) and \(\gamma_{(\text{MC}-\text{BL\_A})}\) relative to the input SNR and bitrate follow similar trends as differential PSNR. Without exception, we
  observed similar trends for all the listed bitrates. The inverse variations of the differential scores with respect to \(\hat{x}_{\text{BL\_A}}\) and \(\hat{x}_{\text{BL\_B}}\) supports our expectation that the proposed postfilter optimally merges information from the two channels to obtain an enhanced multidevice estimate.

  The test was repeated to include reverberation over a range of
  absorption coefficients, \(\alpha=\{0.1, 0.3, ...0.9\}\).
  The results for \(R\in\{16, 32\si{kbps}\}\) are illustrated in
  Fig.~\ref{error_bars}~(b, d).
  While \(\rho_{(\text{MC}-\text{BL\_B})}\) is positive for both
  bitrates over all the listed absorption coefficients, \(\rho_{(\text{MC}-\text{BL\_A})}\) is consistently negative. One reason for this could be that while the
  postfilter  reduces environment noise, as is reflected in the
  improvement with respect to \(\hat{x}_{\text{BL\_B}}\), it may introduce some speech
  distortion, or is unable to completely remove reverberation due to the lack of reverberation model,
  which shows as a drop in the PSNR with respect to \(\hat{x}_{\text{BL\_A}}\). Nevertheless, both \(\gamma_{(\text{MC}-\text{BL\_A})}\) and \(\gamma_{(\text{MC}-\text{BL\_B})}\)
  are positive over both the bitrates and all \(\alpha\), and follow similar variation trends as in the additive noise scenario. Lastly, the positive differential objective scores for both noise types with respect to the MWF indicate that the PSNR and PESQ gains of the proposed postfilter are larger than the gains obtained using the multichannel Wiener filter. This supports our informal observation that Wiener filtering is inefficient in capturing the essential features of speech signals.

  The subjective MUSHRA listening test
  contained eight test items (4 male and 4 female), four of which included
  background noise with reverberation at \(\alpha=0.3\) while the remaining
  items comprised of background noise only at \(\text{SNR}=\SI{15}{dB}\).
  Each test item consisted of five test conditions and the reference clean
  speech signal; a hidden reference and a lower anchor, which was the
  \SI{3.5}{kHz} low-pass version of the reference signal, \(\hat{x}_{\text{MC}}\), \(\hat{x}_{\text{BL\_B}}\), and \(\hat{x}_{\text{BL\_A}}\) were presented as the test conditions; total bitrate was \(R=\SI{32}{kbps}\). As post-screening, we retained the responses from
  only those subjects that rated the hidden reference at more than
  \SI{90}{MUSHRA}~points for all items. Fig.~\ref{mushra} presents the
  consolidated differential MUSHRA, represented as \(\eta\), from 13 participants who passed the
  post-screening; the boxplots show the median and interquartile range of \(\eta\).
  The background noise with reverberation are presented in
  \(\text{items}\,\{1, 2, 3, 4\}\) and the background-noise-only samples are
  \(\text{items}\,\{5, 6, 7, 8\}\). \(\text{Items}\,\{1, 2, 5, 6\}\) are female and the rest are male. \(\eta_{(\text{MC}-\text{BL\_A})}\) was
  positive for all items, indicating that most subjects preferred \(\hat{x}_{\text{MC}}\) over \(\hat{x}_{\text{BL\_A}}\). With
  respect to \(\hat{x}_{\text{BL\_B}}\), the variations were found to be gender dependent.
  While the median \(\eta_{(\text{MC}-\text{BL\_B})}\) points were positive for most male items~(mean-M),
  they were negative for females~(mean-F). Further analysis of the samples revealed that while background noise was
  attenuated in the \(\hat{x}_{\text{MC}}\), speech distortions were introduced into the estimate and
  those distortions were more prominent in the female samples. This problem could potentially
  be addressed by using more informative speech priors, and modifying the signal model to incorporate the effects of
  reverberation.

       \begin{figure}[!t]
          \centering
          \includegraphics[width=0.4\columnwidth]{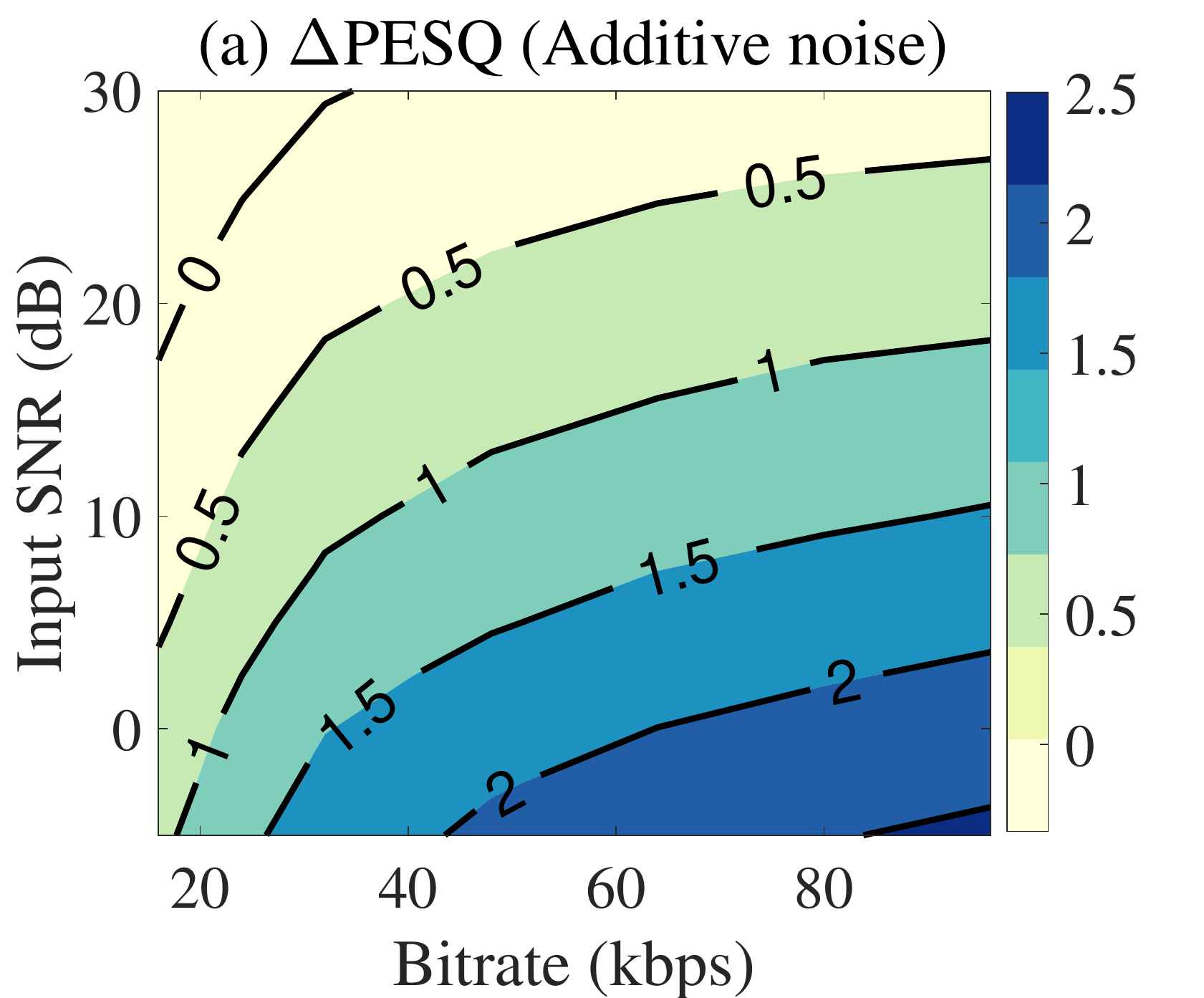}\includegraphics[width=0.4\columnwidth]{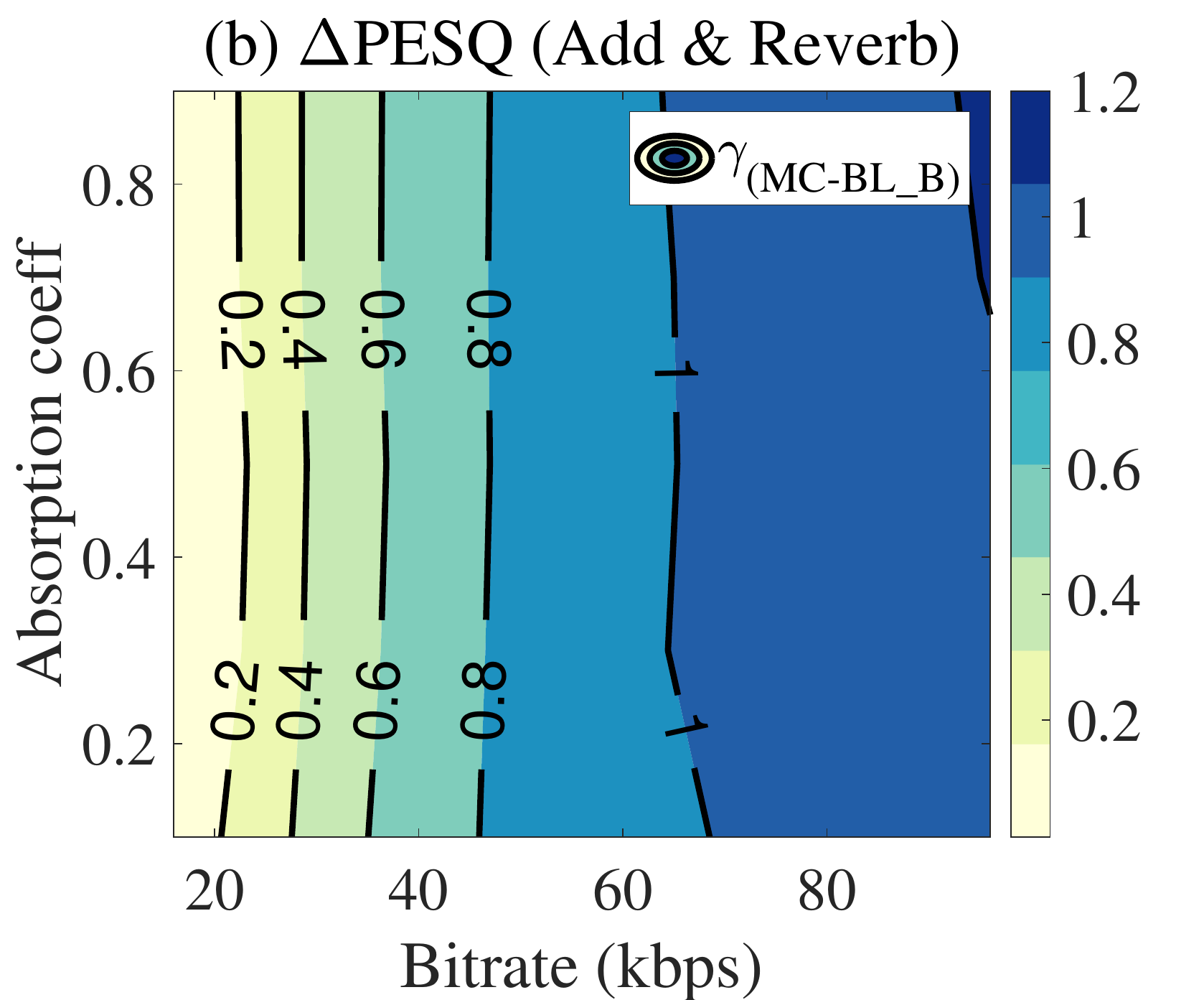}
            \caption{Contour plot showing the differential PESQ,  \(\gamma_{(\text{MC}-\text{BL\_B})}\) jointly over
         bitrates, input SNR and absorption coefficients.}
         \label{contour}
         \end{figure}
  To study the region of optimal performance of the postfilter, we analyzed the
  average \(\gamma_{(\text{MC}-\text{BL\_B})}\) as a function of bitrate and input
  SNRs and absorption coefficient \(\alpha\); the resulting contour plots are
  depicted in Fig.~\ref{contour}. For the additive background noise scenario, the highest gains are at higher bitrates and low input
  SNRs. Furthermore, the negative \(\gamma_{(\text{MC}-\text{BL\_B})}\) over
  \SI{20}{dB} input SNR and below \SI{32}{kbps} implies that the postfilter
  performs sub-optimally in this region; in other words, we gain from a
  multidevice signal estimate when the additive degradation level is below
  \SI{20}{dB} and the total bitrate is greater than \SI{32}{kbps}. In the
  presence of reverberation, we observed that while the total bitrate had an
  impact on \(\gamma_{(\text{MC}-\text{BL\_B})}\), the improvement was fairly
  constant over the range of \(\alpha\) at an arbitrary bitrate, and the
  improvement was positive over the considered input SNR range.
  This implies that the proposed postfilter can also be used to enhance
  signals degraded by reverberation and is not especially sensitive to the
  amount of reverberation, despite the fact that the signal model did not
  explicitly account for distortions from reverberation.

  \section{Conclusion}

  In this work, we proposed a postfilter to enhance speech in an ad-hoc sensor network. The method explored the feasibility
  of using sources degraded by two distinct noise types to obtain an enhanced estimate of the clean speech signal. We demonstrated that by
  distributing the total available bandwidth between two sensors, we can
  achieve signal quality that is higher than a single channel estimate
  operating at full bitrate.
  Further work is needed to address the classic noise-reduction vs. speech distortion
  problem, by incorporating a signal model which takes into account the effects of reverberation, although the objective and subjective results are already encouraging.

\bibliographystyle{unsrt}

\bibliography{references}

\end{document}